\documentclass[preprint,showpacs,preprintnumbers,amsmath,amssymb]{revtex4} 

\usepackage{graphicx}
\usepackage{dcolumn}
\usepackage{bm}
\begin{document}

\title{\bf Microwave absorption/reflection and magneto-transport experiments on high-mobility electron gas. }

\author{S.A. Studenikin$^1$,
M. Potemski$^{1,2}$,  A.S.Sachrajda$^1$, M.Hilke$^3$, L. N. Pfeiffer$^4$, K. W. West$^4$  } 
\affiliation{
$^1$Institute for Microstructural Sciences, NRC, Ottawa, Ontario, K1A OR6, Canada\\
$^2$GHMFL, MPI/FKF and CNRS, BP 166, 38042 Grenoble, Cedex 9, France \\
$^3$Department Of Physics, McGill University, Montreal, Canada H3A 2T8  \\
$^4$ Bell Laboratories, Lucent Technologies, Murray Hill, New Jersey, 07974
}

\date{ September 3, 2004 }   

\begin{abstract}

We have performed simultaneous measurements of microwave absorption/reflection and magneto-transport characteristics of a high mobility two-dimensional electrons in GaAs/AlGaAs heterostructure in regime of Microwave-Induced Resistance Oscillations (MIROs).  
It is shown that the electrodynamic aspect of the problem is important in these experiments.
In the absorption experiments a broad cyclotron resonance line was observed due to a large reflection from the highly conductive electron gas. There were no additional features observed related to absorption at harmonics of the cyclotron resonance. 

In near-field reflection experiments a very different oscillation pattern was revealed when compared to MIROs.  The oscillation pattern observed in the reflection experiments is probably due to plasma effects occurring in a finite-size sample.   The whole microscopic picture of MIROs is more complicated than simply a resonant absorption at harmonics of the cyclotron resonance.  
Nevertheless, the experimental observations are in good agreement with the model by Ryzhii et al. involving the photo-assisted scattering in the presence of a crossed magnetic field and $dc$ bias.
The observed damping factor of MIROs may be attributed to a change in the electron mobility as a function of temperature.   
\end{abstract}
\maketitle
\newpage

\section{Introduction}
It was in 1969 that Ryzhii \cite{Ryzhii1970,Ryzhii1986} theoretically predicted an oscillatory dependence in the photoconductivity of quantized semiconductor films due to the photon-assisted impurity scattering of electrons between Landau levels.  At that time the prediction was not supported by experiments because samples were not available and it did not receive any significant attention.
Much later in 2001, while being unaware of Ryzhii's work, Zudov et al. \cite{Zudov} reported the observation of "Shubnikov -de Haas like" oscillations on a 3$\times$10$^6$cm$^2$/Vs mobility sample with a period corresponding to the cyclotron resonance harmonics, which looked like other oscillatory phenomenon related to Landau quantization.  
Recently, after the observation of states with a vanishing resistance on very high mobility samples $\sim$10$^7$cm$^2$/Vs \cite{Mani,Zudov2}, this phenomenon attracted a lot of interest within the scientific community.  
A large number of theoretical papers have been published to explain this fascinating phenomenon of vanishing resistance (e.g. \cite{Ryzhii1, Vavilov, Durst} and references therein).
In particular, to explain microwave-induced resistance oscillations (MIROs), authors evoke radiation-induced scattering by impurities \cite{Durst,Ryzhii1970} and phonons \cite{Ryzhii1,Ryzhii2}, a non-equilibrium distribution function \cite{Dorozhkin,Dmitriev}, or electron plasma effects \cite{Volkov,Mikhailov} which may lead to an unstable state with a negative conductance and the formation of current domains \cite{Vavilov,Andreev}. 
There are also theoretical models involving the formation of an energy gap at the Fermi level leading to a superconducting state.  \cite{Mani,Fujita}
In addition, as was pointed out by Mikhailov \cite{Mikhailov} the importance of electrodynamic effects - radiative decay, plasma oscillations, and retardation - should not be disregarded in microwave (MW) experiments. This will be the main subject of this paper.

In particular, in this work we examine the electrodynamic issue of the MIROs by performing MW absorption/reflection experiments.
It is shown that the metallic-like reflection by a highly conductive sample is very important, e.g. to explain the broad cyclotron resonance (CR) line in absorption experiments.  
In near-field MW reflection experiments on a small sample the MW detector presents a very different picture in comparison to transport measurements.  Our results are discussed in comparison with existing theoretical models.  

We also address another puzzling feature, pointed out earlier by several groups \cite{Mani,Zudov2}, that MIROs exhibit a much smaller damping with increasing temperature compared to the Shubnikov-de Haas (SdH) effect.  A new approach is suggested to analyze the temperature damping factor of MIROs by fitting the waveform with an exponentially damped sinus function.  Using this approach a reasonable value was obtained for the temperature damping factor of MIROs, which may be attributed to the temperature dependence of the scattering process (electron mobility).

\section{Experimental}
Experiments were performed on a GaAs/Al$_x$Ga$_{1-x}$As heterostructure containing a very high mobility two dimensional electron gas (2DEG) grown by molecular beam epitaxy.   The sample for the transport measurements was cleaved into a rectangular shape of 1.6x5 mm$^2$ with small indium contacts at the edges prepared by annealing in forming gas (mixture of 10\% H$_2$ and 90\% N$_2$) at 420$^\circ$C for 10 minutes. Distance between potential contacts {\it L} was 1.6 mm so that the ratio {\it w/L}=1, where {\it w} is the width of the Hall bar.
 After a brief illumination with a red LED the 2D electron gas attained the following concentration and mobilty n=1.9$\times$10$^{11}$cm$^{-2}$ and  $\mu \simeq$ 8 $\times$10$^6$cm$^2$/Vs at temperature 2 K.
The microwave measurements were performed between 1.4 and 4.2 K in a He$_4$ cryostat equipped with an Oxford superconducting solenoid. 
The magnetic field normal to the sample surface was produced by the superconducting magnet which was carefully calibrated using ESR and Hall effect probes, and the weak antilocalization effect. \cite{ourWAL}  
An Anritsu, model 69377B, signal generator with a typical output power of a few milliwatts and frequencies up to 50 GHz was used as the source of microwaves.  
The longitudinal and transverse resistances of the 2DEG were measured using standard {\it ac} techniques on an AVS-47 resistance bridge.
To obtain direct information concerning the modulation of the electron density of states and the shape of the Landau levels, low temperature investigations of the Shubnikov-de Haas (SdH) oscillations were performed on a He$_3$/He$_4$ dilution refrigerator at temperatures down to 30 mK.

Figure 1 shows schematic diagrams of the reflection (a) and absorption (b) experiments, correspondingly.
The microwave radiation was delivered into the cryostat using a copper-beryllium semirigid, 0.085" diameter, coaxial cable. 
\begin{figure}[tbp]
\includegraphics[width=68mm,clip=false]{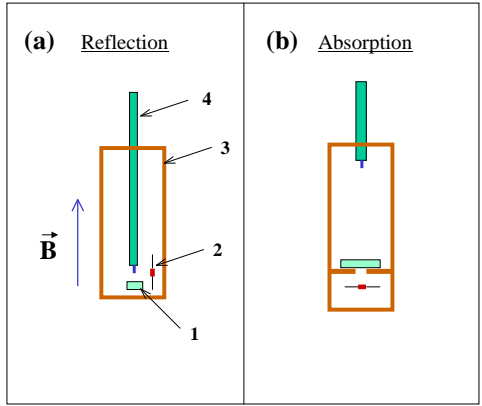}
\caption{ Diagrams of the reflection (b) and absorption (b) experiments. 
1 - sample, 2 - small microwave sensor made of a carbon thermoresistor, 3 - an absorbing cavity made of a MW absorbing material, 4 - copper beryllium semirigid coaxial cable.} 
\label{fig1}
\end{figure}
The coaxial cable (4) was terminated with an antenna, the central conductor extending a few millimeters from the outer jacket. This antenna was used to irradiate the sample with MWs by placing the antenna a few millimeters above the sample in reflection and MIROs experiments, and a few centimeters away in the absorption experiment.   
A small cylindrical absorbing cavity (3) made from a MW absorbing material was constructed around the sample (1) to suppress cavity modes which could potentially arise from the metallic components of the cryostat.  

The MW reflection experiment (Fig. 1(a)) was performed in a near-field geometry on a small sample under strong excitation power (in regime of MIROs) when pronounced oscillations in the magnetoresistance  were observed and the measured resistance could go to zero.   To monitor the MW power during the MIROs, the MW sensor made of an Allen Bradley carbon thermo-resistor was placed close to the sample, sideway above it. 
In order to reduce heat exchange with the helium bath the thermoresistor was encapsulated in a small plastic housing, transparent for MWs. 
It should be noted that this was a rather qualitative experiment since the MW field distribution was not very well defined because of a small size of the sample and the near field geometry, Fig.~1~(a).   It was impossible, therefore, to carry out quantitative calculations of the MW field distribution; however, this experiment still provided us with useful information for the conditions in which most of MIROs experiments were carried out.   

For the absorption experiment we took special care for the geometry so that a quantitative description could be made.  
A much larger piece of sample $\sim$9$\times$9 mm$^2$ was used in this experiment in order to exclude the magneto-plasmon shift of the cyclotron resonance (CR) due to a finite size of the sample. In addition, a mask made of the MW absorbing material with a $\sim$3 mm diameter hole was placed before the sensor to eliminate possible edge effects and ensure that only the radiation passing through the central part of the sample reached the MW sensor (Fig. 1(b)). 
In the absorption experiment the antenna was placed about 30 mm away from the sample to ensure the small-signal regime and a transverse geometry for the electromagnetic field. 
In this experiment our intention was to quantitatively study the MW absorption by a high mobility 2DEG and search for evidence of possible absorption at harmonics of the CR $\omega = j\omega_c$  ({\it j}=1,2,3...) or any other resonances related to the MIROs.
%
\begin{figure}[tbp]
\includegraphics[width=68mm,clip=false]{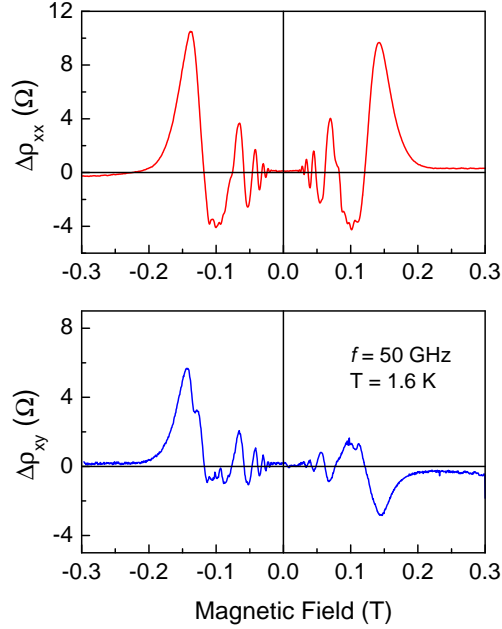}
\caption{ Experimental curves of $\Delta \rho_{xx}$ {\bf (a)} and $\Delta \rho_{xy}$ {\bf (b)} obtained as a difference between traces with and without MW radiation ({\it f}= 50 GHz) on a high-mobility   GaAs/AlGaAs sample for both directions of the magnetic field.  \\ } 
\label{fig2}
\end{figure}
Let us begin with an example of MIROs shown in Fig. 2 (a) on a sample irradiated with {\it f}= 50 GHz.  In this figure $\Delta \rho_{xx}$ is the experimental dependence obtained as a difference between two identical B-sweeps with and without MW irradiation.  In accordance with observations by other authors \cite{Mani,Zudov2}, the oscillations  are periodic as a function of the inverse magnetic field, and there is no resistance change at the exact points of the cyclotron resonance and harmonics:  
$\omega / \omega _c = j$, where $j$=1,2,3... is an integer number, $\omega = 2\pi f$ is the microwave circular frequency, $\omega _c = eB/m^*$ is the cyclotron frequency, $m^*$ is the effective electron mass. \cite{Vavilov,Durst}  
In spite of the large resistance oscillations leading to an apparent zero resistance state, the off-diagonal (Hall) component of the resistivity  did not reveal noticeable oscillations in the first experiments (see e.g. refs. \onlinecite{Mani,Zudov2}).  After a more careful examination it was discovered, however, that the off-diagonal component did contain oscillations as well. \cite{ourMW, ManiRxy}.  Figure 2 (b) shows oscillations of $\Delta \rho_{xy}$ from the same sample.  
As expected, the MW induced oscillations in the Hall effect are an anti-symmetric function in magnetic field. It is shown that oscillations in $\rho_{xy}$ cannot be attributed to an admixture of the $\rho_{xx}$ component.\cite{ourMW}  In the experiment shown in Fig. 2(b) the  oscillation amplitude of $\Delta \rho_{xy}$  was approximately half that of $\Delta \rho_{xx}$ in Fig. 2(a).  It is in qualitative agreement with existing theoretical models which predict comparable values for the amplitude of oscillations in $\rho_{xx}$ and $\rho_{xy}$ \cite{Durst, RyzhiiRxy}.

\subsection{The absorption experiments}
In this experiment our aim was to quantitatively study MW absorption by a high mobility 2DEG and search for evidence of absorption at harmonics of the CR ($\omega = j\omega_c$  with {\it j}=1,2,3...) or any other resonances related to the MIROs.  Geometry of this experiment is shown in Fig. 1(b). 
It was not a straightforward task to perform this experiment quantitatively and to achieve the expected result for the classical CR. As described above, special care was taken to ensure a far-field geometry and to exclude plasmon and edge effects.  Note,  the transmission was measured in this experiment, which was proportional to absorption.  The measurements were taken in arbitrary units.
As it was mentioned above, for the absorption experiment we used a larger piece of a high-mobility sample with $\sim$9x9 mm$^2$ dimensions from the same source (Lucent Labs).   This sample had an electron concentration of 1.8x10$^{11}$ cm$^{-2}$, and a mobility of 1.4$\times$10$^6$ cm$^2$/Vs which did not change after illumination. 

Figure 3 shows the experimental dependences  of the carbon thermoresistor response R$_t$ for ({\it f}=49.7 GHz) along with the simulations.  Simulations showed that absorption by 2DEG was proportional to the transmission measured by the MW sensor in relative units.
Top curve (1) in Fig. 3 is a "dark" magnetoresistance trace (no MW irradiation) of the carbon thermoresistor.  Note that there was a small, less than 0.1\%, non-monotonic magnetoresistance in R$_t$, which had to be taken into account in the small signal regime where the signals were small and comparable to the dark magnetoresistance.  
After irradiation with microwaves the whole R$_t$ curve shifted toward  smaller resistances due to the heating by the transmitted radiation through the 2DEG sample.  In addition, under the MW radiation, two peaks are evident in the thermoresistor response (curve 2 in Fig.~3) at the exact positions of the CR $\omega_c = \pm \omega$.   
Magnetoplasmon shift of the CR is not noticeable here because of a larger size of the sample. \cite{Klitzing1993,Pan2003,Kukushkin2003} 

Due to the small signals and high sensitivity of the thermoresistor (note the scale mark in Fig. 3), the absorption experiment had to be done at 4.2 K, because temperature controlling resulted in large oscillating drifts during the magnetic field sweeps.  Even at 4.2 K, a noticeable temperature drift occured under the MW radiation that is evident from a slight asymmetry of the $\Delta R_t$ curve in Fig. 3.  Relative transmission response $\Delta R_t$ (curve 3) in Fig.~3, which is proportional to the absorption by 2DEG, was obtained by substracting the background trace 1$^\prime$ from the R$_t$ response with MWs (curve 2).
It is evident from Fig.~3 that the central peak around B=0  is an artifact related to a B-dependent sensitivity of the MW sensor at 4.2 K and, therefore, should be disregarded.  

The small magnetoresistance effect of the sensor was negligible in the reflection experiments (large signal regime) at lower temperatures, which will be discussed later. 
\begin{figure}[tbp]
\includegraphics[width=68mm,clip=false]{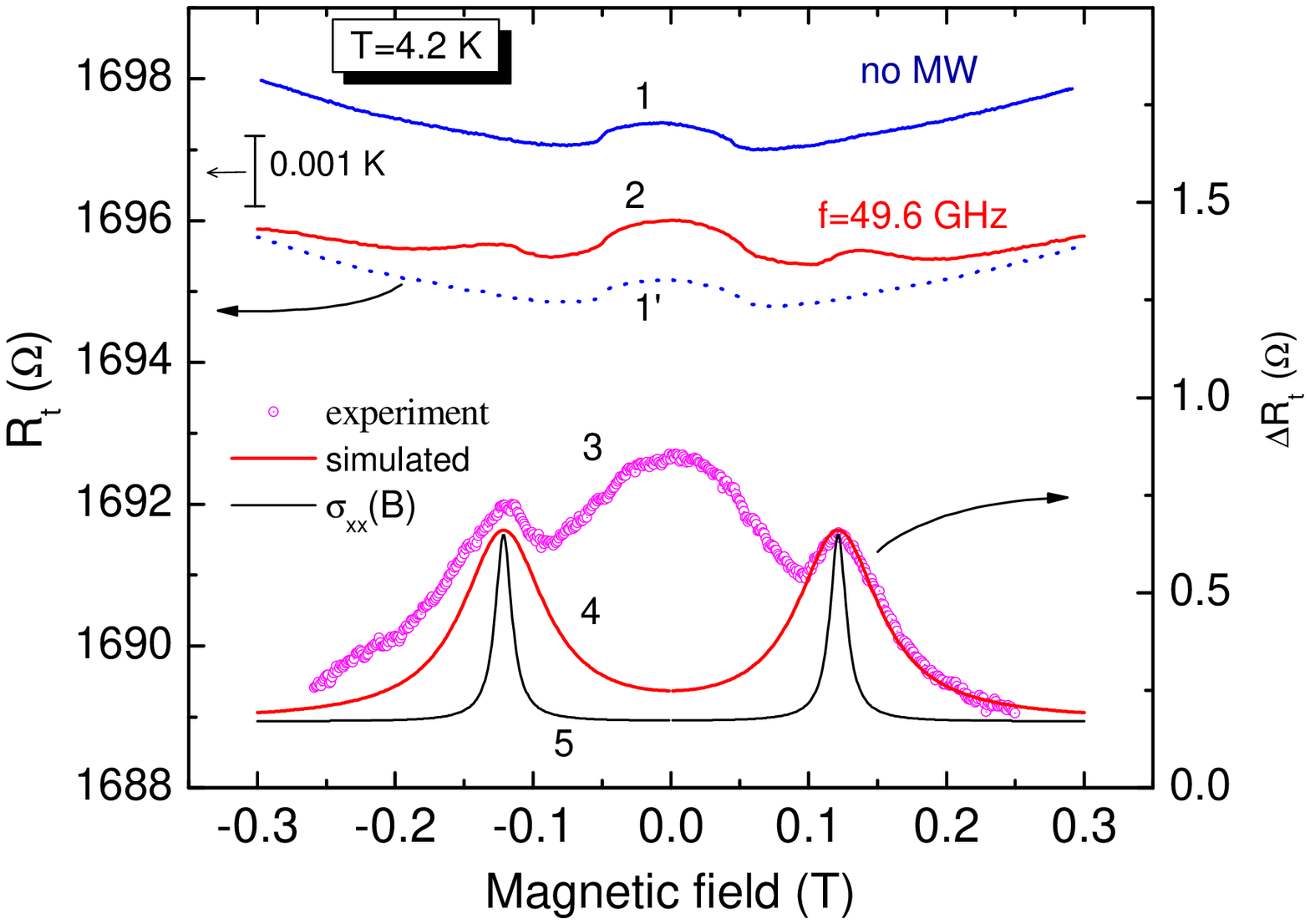}
\caption{Absorption by a high mobility 2DEG vs magnetic field at {\it f}=49.7 GHz, and T=4.2 K. \\
Lines 1 and 2 are magnetoresistance of the thermo-resistor (Fig. 1 (b)) without and with microwaves, correspondingly; 
curve 3 is a relative change of the MW sensor response due to the transmitted MW radiation obtained by subtracting background dependence (curve 1');  curve 4  is the theoretical dependence from Eq.~(2);  curve 5  is the real part of the dynamic conductivity $\sigma_{xx}(\omega)$ by Eq.~(1) in arbitrary units.
 } 
\label{fig3}
\end{figure}
Solid line (5) in Fig. 3 is the real part of the dynamic conductivity in arbitrary units calculated using the Drude expression:
\begin{equation}
\sigma ^{\pm}(\omega,B)=\frac{e n \mu}{1-i(\omega \pm \omega_c)\tau}
\label{eq1}
\end{equation}
where {\it e} is the electronic charge, $\tau$ is the transport relaxation time, $\mu=e\tau/m^*$ is the electron mobility, {\it n} is the sheet concentration of 2DEG,  $\omega$ and $\omega_c$ are the MW and cyclotron cyclic frequencies respectively, $i=\sqrt{-1}$ is the imaginary unit.

It is seen in Fig. 3 that the Drude conductivity gives much narrower peaks (dashed line) of the CR that would be the case for a low conductivity sample.   
In real situation of highly conductive samples used in MIROs experiments, it is important to consider the electrodynamic aspect of the problem based on the Maxwell's equations. \cite{ourMW2}  The absorbed power is calculated as 
$\Delta P_{\text{a}}$=$<$Re({\it \^E$^{*}$}$\sigma(\omega)${\it \^E)}$>$, where {\it \^E} is the electric component of the MW field in the plane of the 2DEG.  
The absorbed power $\Delta P_{\text{a}}/P_i$  for linear polarized radiation is given by the following expression \cite{ourMW2}:   
\begin{equation}
	\frac{\Delta P_{\text{a}}}{P_i} =\sum_{\pm} \frac{Re(\bar \sigma^{\pm})}{(1+\kappa+Re(\bar \sigma^{\pm}))^2+(Im(\bar \sigma^{\pm}))^2}
	\label{eq2}
\end{equation}
where  $\kappa$ is the refraction index, which was 3.6 for GaAs in our calculations,  
$\bar \sigma^{\pm} = \sigma^{\pm} Z_0$ is the normalized sheet conductivity of the 2DEG, and  $Z_0=\sqrt{\frac{\mu_0}{\epsilon_0}}$=377 $\Omega$ is the impedance of free space.
Higher order corrections due to interference effects were not taken into account in Eq.~(2).

Solid line (4) in Fig. 3 is the theoretical curve from Eq.(2) using the transport parameters for this sample (n=1.8x10$^{11}$ cm$^{-2}$, and $\mu$= 1.4$\times$10$^6$ cm$^2$/Vs). 
It is evident from Fig. 3 that the simulation (solid line) describes the experimentally observed CR peaks quite well. The calculated absorbed power at the CR peak in this sample was small, about 3 \%,  due to the reflection. In the higher mobility samples used for the reflection and MIROs experiments below the resonance absorption was even smaller, less than 1 \%. 
From numerical calculations of Eq.(2) it follows that the maximum absorption occurs at $\sigma_0 \simeq 4/Z_0$ and the absorption decreases with increasing conductivity when $\sigma_0 \geq 4/Z_0$.  

In the absorption experiment (Fig. 3) in the small signal regime we did not observe any features due to harmonics of the CR (within an accuracy of a few percents).  Our simulations indicate that if a small resonant absorption existed at the CR harmonics, e.g. at $\omega = j\omega_c$ with {\it j}=2,3...  it would produce sharp peaks which would be easily noticeable on the experimental curve.  Broadening would not be important for harmonics because of the small value of the partial {\it ac} conductivity $\sigma_{j\omega}(\omega)$  which would cause a much smaller broadening effect if $\sigma_{j\omega}(\omega) < 4/Z_0$. For CR harmonics the width would be close to that defined by the electron mobility FWHM$\simeq$2/$\mu$ in Eq.~(1).

\subsection{The reflection experiments}
MIROs experiments usually have a complicated and not well defined  geometry of the electromagnetic field.  In particular, the sample is small and the exact direction and space configuration of the MW field is not well defined.  Therefore, Eq.(2) cannot be applied to describe the absorption by high-mobility 2DEG samples in MIRO experiments.

Recently, it was emphasized by Mikhailov  \cite{Mikhailov2}, that there are other electrodynamic effects to be considered, which are radiative decay, plasma oscillations, and retardation (see elso ref. \onlinecite{Kukushkin2003}). 

Despite these complications, our intention was to examine what was happening with the MW field in the space around the sample during MIROs at high excitation powers with the antenna placed close to the sample. A small MW sensor made of a carbon thermo resistor was placed sideways above the sample (Fig.~1 (a)).  The geometry resembled that for reflection experiments in optics, but in fact it was more accurately a near-field geometry due to the much larger wavelength of the MW radiation. 
In this way we measured the MW intensity variation near the sample during the magnetic field sweeps.

Figures 4 (a) and (b) show examples of the thermoresistor response ($R_t$, left axis) at different MW powers along with the simultaneously measured MIROs ($R_{xx}$, right scale) for {\it f}= 49.75 GHz and f=38.97 GHz.   
Data in Fig. 4 (a) and (b) were taken at a slightly different temperatures, at 1.38 and 1.47 K correspondingly, which did not affect the oscillation pattern but caused  resistance change of the thermo-resistor $R_t$ because of its high T-sensitivity.
\begin{figure}[tbp]
\includegraphics[width=68mm,clip=false]{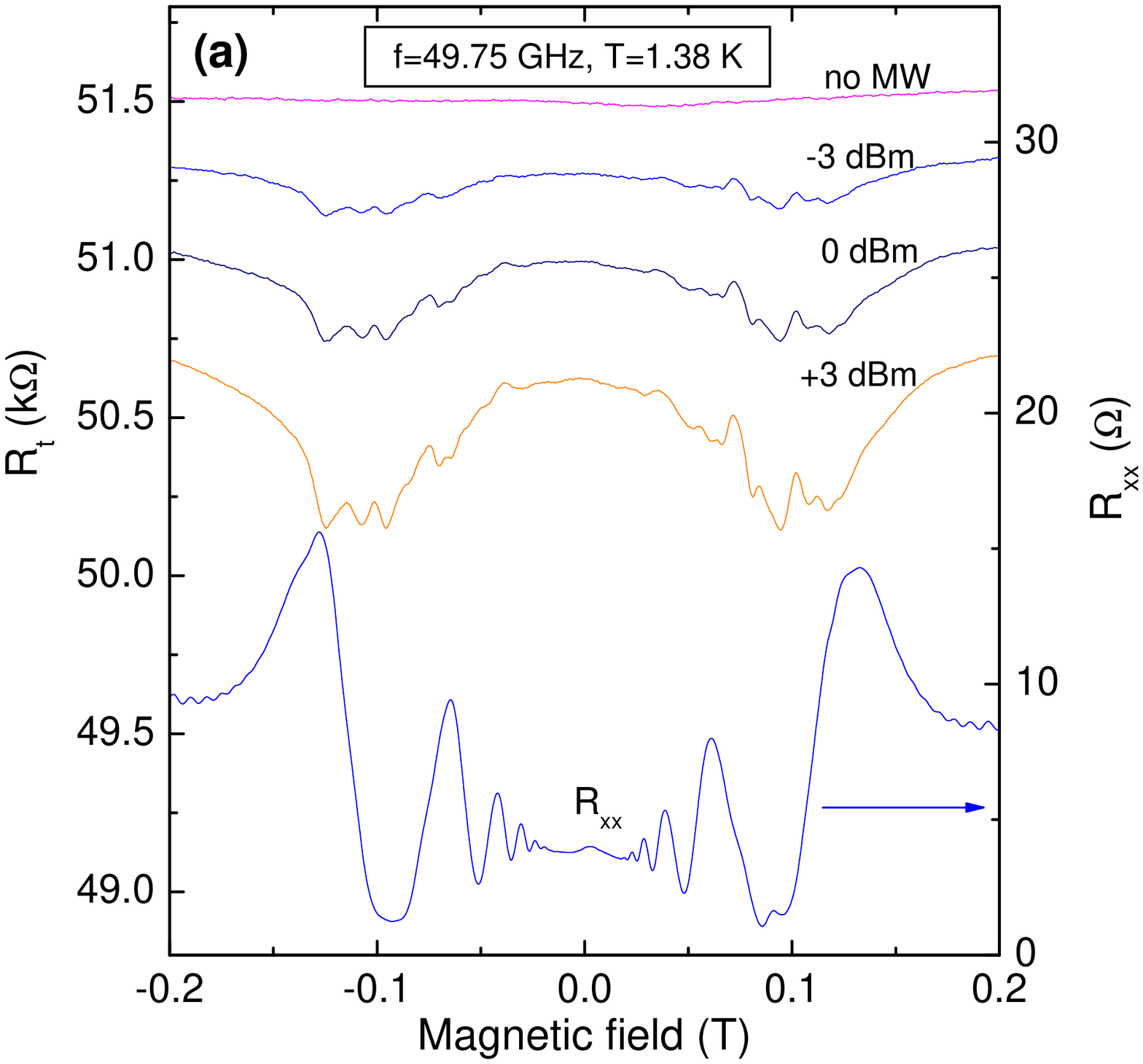}
\includegraphics[width=68mm,clip=false]{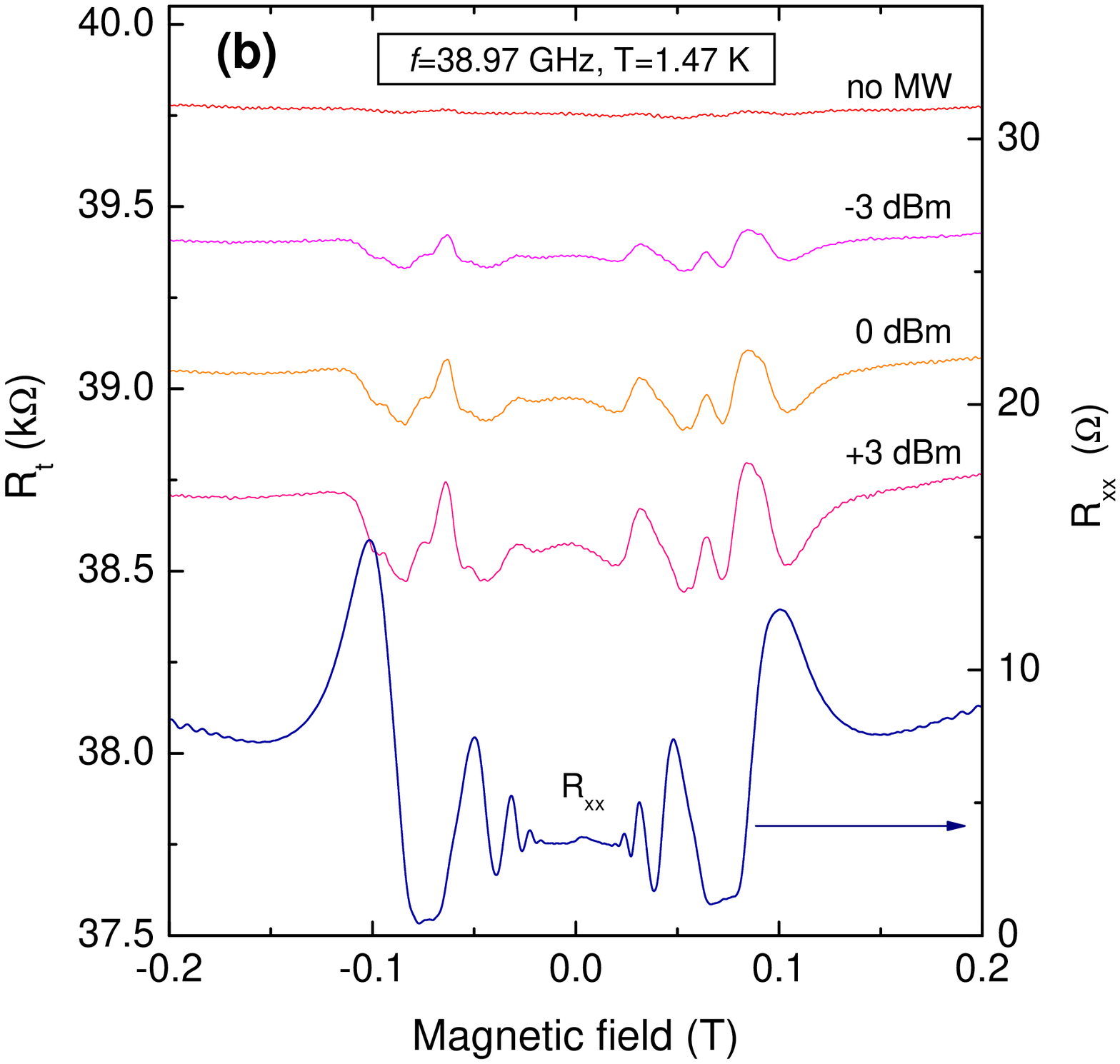}
\caption{ Respose of the MW detector placed close to the sample in the MIROs regime at different powers (right axis) and the corresponding MIROs trace (bottom trace, left axis).
(a) {\it f}=39.75 GHz, T= 1.38 K;  (b) {\it f}=38.94 GHz, T= 1.47 K.
 } 
\label{fig4}
\end{figure}
It can be seen in Figs. 4 (a) and (b) that the oscillation pattern of the MW field detected by the thermoresistor is very different from the MIROs (the bottom trace in both figures).  
The MW sensor response is flat for magnetic fields larger than the CR field $\omega_c > \omega$, indicating that the MW field remained relatively constant at large fields.  In magnetic fields around and below the CR ($\omega_c \leq \omega$) oscillations are visible in the $R_t$ response, whose position and waveform does not depend on the MW power or temperature, but does depend on the MW frequency and the illumination used to change the electron density (not shown here).  
It should be noted that at smaller magnetic fields, where the MIROs cease, the MW response of the thermoresistor also stops to oscillate.  This strongly suggests that the MIROs and the MW field oscillations are related, while revealing different patterns.

It is important to note that the oscillations of the MW sensor $R_t$ occur continuously throughout the whole MIROs regime without any particular changes within the flat areas in $R_{xx}$, the so-called zero-resistance states.  If the zero-resistance states were due to a superconductivity effect \cite{Mani,Fujita}, one would have to observe a dramatic change in the reflectivity. This is not observed in Fig.4.

Another proposed explanation for the zero-resistance state employs the formation of current domains due to a negative conductivity. \cite{Vavilov,Andreev}
While domain formation maybe an important element in the explanation of the apparent zero-resistance state it is clear that the domains do not reveal themselves in the reflection since, as mentioned, $R_t$ oscillates throughout the whole MIROs regime and not only in regions of the zero-resistance states.   
It is likely, that the MW field  oscillations ($R_t$ in Fig.4) are caused by bulk and edge plasmons  due to the finite size of the sample.  \cite{Kukushkin2003,Mikhailov2}  Further experiments are being planned to clarify this.

At this stage, it can be concluded that processes seen in transport and reflection experiments are different.  It should be noticed that sometimes a fine oscillation pattern can be observed in $R_{xx}$ on top of the MIROs, e.g. it is evident in Fig. 2.  These smaller oscillations may be attributed to the magneto plasmon effects, which are strong in $ac$ reflection experiments but are only a second order correction in transport.
The observations in the MW absorption/reflection experiments discussed above are in agreement with the model of MIROs based on non-resonant absorption between Landau levels.  \cite{Ryzhii1970, Ryzhii1986,Durst, Dmitriev, Dmitriev2} 

In this picture, MW radiation results in a large number of single-particle electron transitions.  On our opinion, these electron transitions occur through a strong electrodynamic interaction of electrons with all kinds of bulk and edge plasmons. Most of the electron transitions between Landau levels are non-resonant. Evidence for this comes from a broad cyclotron line and absence of sharp resonant transitions at CR harmonics in Fig.~3. 
 
In MIROs experiments there is a strong modulation in the electron density of states \cite{ourMW2}.  In this case the non-resonant electron transitions between Landau levels ($\Delta N \geq$ 2) are enhanced by involving a third quasi-particle into the scattering mechanism, e.g. impurities or phonons. \cite{Ryzhii1, Vavilov,Durst}
Most of the electron transitions do not result in changes to transport, since, to first order, the conductivity of the degenerate electron gas does not depend on temperature.

On the other hand, there are particular transitions which are very important for the measured resistance because they induce changes in the distribution function in $\vec k$ space by forcing electrons to predominantly scatter in a particular direction along or against the $dc$ bias.  The direction of the predominant scattering depends on the detuning conditions away from the cyclotron resonance and its harmonics.   In other words, the degenerate 2DEG system serves as an extremely selective sensor only for those photon-assisted electron transitions which are pushing electrons up or down hill (along or against the bias).   

The elegance of this effect lies in the fact that, though a MW photon has negligible momentum $k_{MW}$, electrons effectively attain a very large momentum of the order of $k_F$ during each scattering event, therefore, under certain conditions this mechanism  can result in a very large effect, specifically on very high-mobility samples at $\hbar \omega = (j \pm 1/4)\hbar \omega_c$.  This mechanism requires a $dc$ bias which transforms it to a resistance change, but it may also be considered as a light induced drift effect.  

In other words, it is possible to classify MIROs as belonging to a broader class of phenomena, the well-known light-induced drift (LID) of electrons in semiconductors or molecules in gases.\cite{Shalagin,Skok,Shalaev} 
For both the radiation produces non-symmetric electron transitions in velocity (momentum) space leading to a non-symmetric velocity distribution.  In both cases the effect change the sign while sweeping trough the resonance $\sim (\omega - \omega_0)$, and the effects are zero at exact resonance $(\omega = \omega_0)$.
In both cases, MIROs and LID, light affects the electron distribution function with a corresponding momentum of the order of a typical electron thermal or Fermi momentum.  In each absorption event electron gains large momentum, much larger than the photon momentum, therefore the radiation-induced currents (drift in gases) is much larger than that produced by the radiation pressure.  
The "amplification" factor is about $k_F/k_{MW} = \lambda_{MW}\sqrt{2\pi n}$ which in case of MIROs gives us a number of about $\sim 10^6$.

There are differences, however, between MIROs and LID at the microscopic mechanism level.
In LID in gases the asymmetry of the excitation in momentum space $\vec k$ arises from the Doppler shift, because particles moving along or against the incident radiation wave vector have slightly different excitation energies due to the Doppler effect.\cite{Shalagin, Skok}  
In semiconductors, due to a more complicated energy band diagram, there are many other mechanisms which lead to a light-induced drift detected by the current or a voltage.\cite{Shalaev}
In this sense, the mechanism that results in MIROs discussed above is just a new one.

In LID the drift direction is defined by the light momentum.
In MIROs the direction is defined by the $dc$ bias, not by the radiation, and the resulting diffusive current arises from a combined effect of $dc$ electric field, strong modulation of the density of states and the microwave radiation.  In a magnetic field a periodical modulation of the density of states exists and therefore MIROs become a periodic effect as opposed to LID originating from a single resonance.

\subsection{Temperature damping of MIROs}
Another particular feature of MIROs is that they persist up to much higher temperatures than Shubnikov-de Haas (SdH) oscillations. The activation energies extracted from the individual zero-state minimum were found to be in a range of 10 to 20 K \cite{Mani,Zudov2}, several times larger than either the MW photon or cyclotron energies.  
\begin{figure}[tbp]
\includegraphics[width=68mm,clip=false]{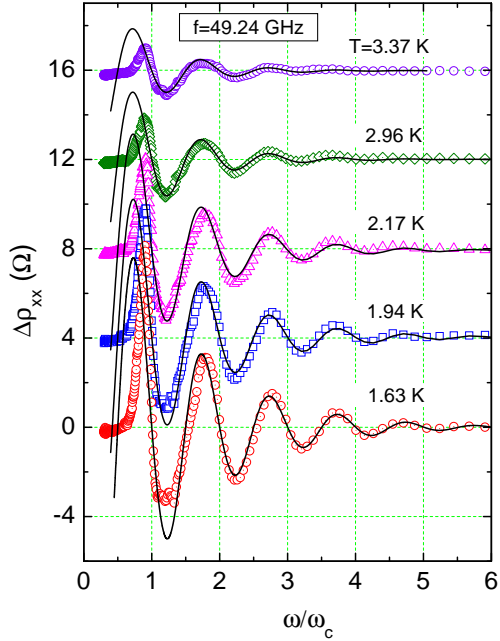}
\caption{ Microwave-induced oscillations in the resistivity of the high-mobility sample at different temperatures.  Solid lines are the best fits with a damped sinus function (Eq.~ 3)}  
\label{fig5}
\end{figure}
Figure 5 shows traces of MIROs for f=49.2 GHz at different temperatures as a function of the normalized inverse magnetic field $\omega / \omega _c$.   
To estimate the activation energy we suggest an analysis of the oscillation amplitude through fitting the experimental curves with a  damped sinus function:
\begin{equation}
\Delta\rho_{xx} = - A \exp(-D_{M}/\hbar\omega_{c}) \sin(2\pi\omega/ \omega_{c}),
\label{eq4}
\end{equation}
where $D_M$ is the temperature damping factor, {\it A} is a positive coefficient.
All curves in Fig. 5 are fitted with a damped sinus function.
It is seen that Eq.~(3) describes MIROs quite accurately except for the zero-resistance state region where the experimental dependence of $\rho_{xx}(B)$ becomes flat rather than following the sinus wave into the negative region.

 The experimental maximum positions for the first three peaks in Fig. 5 are noticeably shifted towards the CR resonance and its harmonics ($\omega / \omega _c$=1,2 and 3) when compared to the minima/maxima of the sinus function.  This effect is attributed to a strong modulation in the density of states. \cite{Zudov3}  
\begin{figure}[tbp]
\includegraphics[width=68mm,clip=false]{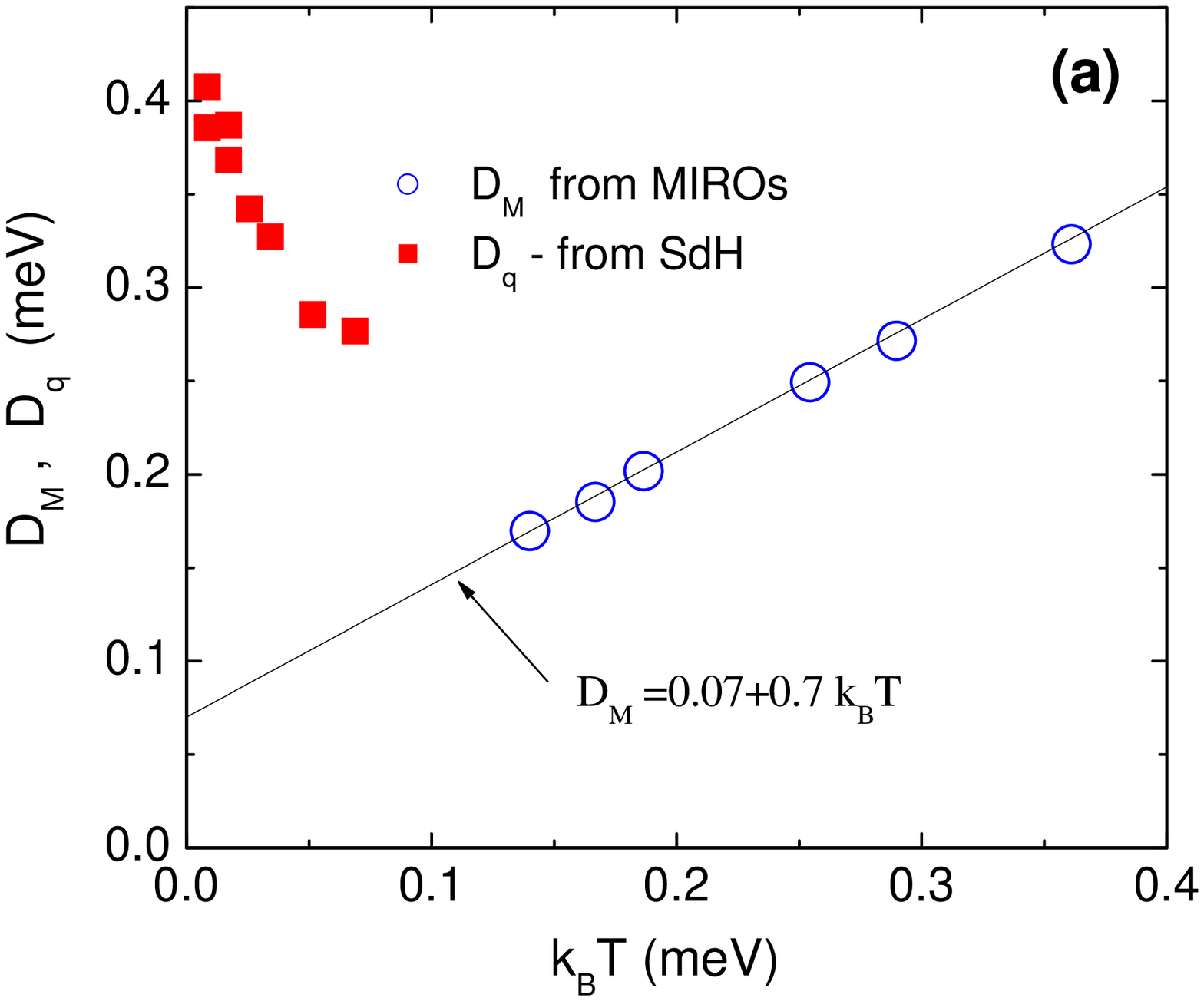}
\includegraphics[width=68mm,clip=false]{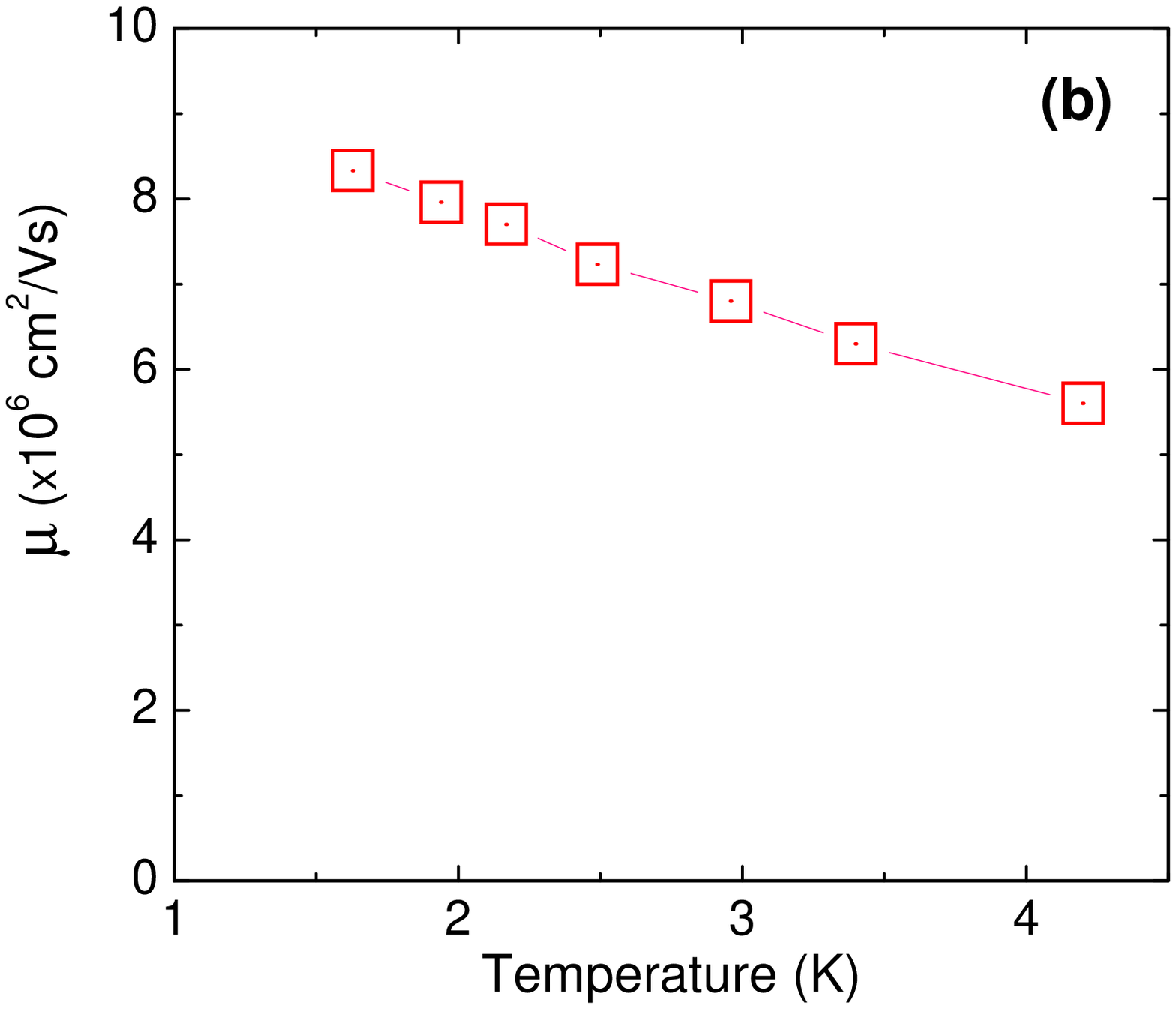}
\caption{ {\bf (a)} Temperature damping  factor $D_M$ of the microwave oscillations obtained from the fits to the experimental curves in Fig. 5 (open circles), and quantum damping parameter from low-temperature SdH measurements.   \\
{\bf (b)} Temperature dependence of the electron mobility in the same temperature range.}
\label{fig6}
\end{figure}
The damping parameters obtained from the fits in Fig.~5 are plotted in Fig.~6 (a) as a function of the temperature in meV units. 
The MIROs damping factor $D_M$ followed a linear function $D_M$=0.07+0.7$k_BT$, which is $\sim k_BT$ expected for a two-level system where the process should depend on the difference between emitted and absorbed MW photons.\cite{Ryzhii3}      
Therefore, using the new approach we obtained reasonable values for the damping coefficient $D_M \sim k_BT$, which did not exceed the characteristic parameters of the experiment - temperature, microwave, and cyclotron energies.

Durst et. al. \cite{Durst} and Lei et. al. \cite{Lei} predicted that the whole temperature dependence of the MIROs should come merely from the T-dependence of the electron scattering process, which can be judged by the electron mobility.
Figure 6 (b) shows the change in the electron mobility for the same set of data.  It is evident that mobility had a similar change as the damping factor $D_M$ in Fig.~6~(a).  Therefore, it is important to consider the temperature change in the mobility, when analyzing the damping/activation behaviour of the MIROs.
A detailed study of the damping factor and MIROs waveform will be published elsewhere.

We also conducted transport measurements on the same sample after a similar illumination at low temperatures (down to 30 mK) in order to compare damping parameter from MIROs with the quantum damping parameter which controls the SdH effect.\cite{ourMW2}
In Fig. 6 solid squares present the quantum damping parameter obtained from the SdH oscillations at low temperatures.  
   The quantum damping parameter, $D_q$ is frequently referred also as a Dingle temperature $T_q$ with $D_q=2\pi^2k_BT_q$.
It is seen from Fig. 6 (a) that $D_q$ is not a constant value $vs$ temperature. The electron transport mobility did not change in the temperature range between 30 and 600 mK, so one would expect that $D_q$  would also not change. 
Although $D_q$ depended somewhat on the temperature, it was still several times larger than the value of $D_M$ obtained from MIROs by extrapolating the dependence in Fig.~6~(a) to zero temperature.
A qualitative explanation of the difference between $D_q$ and $D_M(0)$   can be attributed to the different microscopic origin of these effects.  MIROs are a quasi classical phenomenon requiring only a modulation of the density of states, while SdH is essentially a quantum effect depending also on the Fermi energy.  Therefore, even small fluctuations of the electron density over the sample surface may result in an additional damping mechanism increasing $D_q$ for the SdH effect.

\section{Conclussion}
We have experimentally studied microwave absorption and near-field reflection under the conditions for which MIROs are observed on a high mobility GaAs/AlGaAs 2DEG sample.  
It is shown that electrodynamic effects are important in these experiments.

In the absorption experiment a broad CR line was observed due to the large reflection by the highly conductive 2DEG. There were no features observed related to absorption at the harmonics of the cyclotron resonance. 

Near-field reflection experiments revealed a very different oscillation pattern when compared to MIROs.  The oscillation pattern observed in the reflection experiment could not be explained by models based on superconductivity or domain formation but was rather related to plasma effects in a confined sample.   The whole microscopic picture of MIROs is more complicated than simply resonant absorption at harmonics of the cyclotron resonance.
Nevertheless, the experimental observations including the temperature damping are in a good agreement with the model by Ryzhii et al. and Durst et al. of photo-assisted scattering in the presence of a crossed magnetic field  and $dc$ bias.
In this scenario, MIROs could be classified as belonging to a broader class of light-induced drift effects originated from an asymmetric excitation of electrons in momentum space. 
 
\subsection*{Acknowledgements}
We thank D.G. Austing and P.T. Coleridge for the interest in this work and for useful comments.
S.A.S and A.S. also acknowledge support of The Canadian Institute for Advanced Research (CIAR).  S.A.S. appreciates useful discussions and communications with V. Ryzhii, R. Mani, S. Mikhailov, and A. Govorov.

\end{document}